\documentclass[lettersize,journal]{IEEEtran}
\usepackage{amsmath,amsfonts}
\usepackage{algorithmic}
\usepackage{algorithm}
\usepackage{array}
\usepackage[caption=false,font=normalsize,labelfont=sf,textfont=sf]{subfig}
\usepackage{textcomp}
\usepackage{stfloats}
\usepackage{url}
\usepackage{verbatim}
\usepackage{graphicx}
\usepackage{cite}
\newcommand{\Define}{\stackrel{\Delta}{=}}

\begin{document}

\title{A Novel Precoder for Peak-to-Average Power Ratio Reduction in OTFS Systems}
\author{Saurabh Prakash, Venkatesh Khammammetti and Saif Khan Mohammed, ~\IEEEmembership{Senior Member,~IEEE}
\thanks{Saurabh Prakash and Saif Khan Mohammed are with the Department of Electrical Engineering, Indian Institute of Technology Delhi, India (E-mail: Saurabh.Prakash@ee.iitd.ac.in, saifkmohammed@gmail.com). S. K. Mohammed is also associated with Bharti School of Telecom. Technology and Management (BSTTM), IIT Delhi. Venkatesh Khammammetti is with the Department of Electrical and Computer Engineering, Duke University, USA (E-mail: venkatesh.khammammetti@gmail.com). The work of S. K. Mohammed was supported by the Jai Gupta Chair at I.I.T. Delhi.}

\thanks{This work has been submitted to the IEEE for possible publication. Copyright may be transferred without notice, after which this version may no longer be accessible.}}



\maketitle

\begin{abstract}
We consider the issue of high peak-to-average-power ratio (PAPR) of Orthogonal time frequency space (OTFS) modulated signals. This paper proposes a low-complexity novel iterative PAPR reduction method which
achieves a PAPR reduction of roughly $5$ dB when compared to a OTFS modulated signal without any PAPR compensation. Simulations reveal that the PAPR achieved by the proposed method is significantly \emph{better} than that achieved by other state-of-art methods. Simulations also reveal that the error rate performance of OTFS based systems with the proposed PAPR reduction is similar to that achieved with the other state-of-art methods.

\end{abstract}


\section{Introduction}
Next generation communication systems are envisaged to support high throughput and reliable communication even for high mobility scenarios (e.g., aircraft/UAV communications, high speed train, etc) \cite{ITU2030}. The performance of Orthogonal Frequency Division Multiplexing (OFDM) based systems used in 4G/5G systems is known to degrade severely in high mobility scenarios due to the high channel Doppler shift \cite{Wang2006}. 

Recently introduced Orthogonal Time Frequency Space (OTFS) modulation has been shown to be robust to mobility induced Doppler spread \cite{otfsorg, Hadani2018, SKM_OTFS, OTFSoverview}. In OTFS modulation,
information symbols are embedded in the delay-Doppler (DD) domain and therefore detection \cite{Raviteja2018, SLi2022, VarBayes2020} and channel estimation \cite{OTFSEst1, OTFSEst2, OTFSEst3, SKM_EST } are also carried out in the DD domain. A collection of literature on OTFS is available in \cite{Bestreadings2022}. 

In an OTFS transmitter, the information symbols are first converted to time-frequency (TF) domain symbols which are then converted to time-domain (TD) transmit signal using an OFDM modulator. Due to this transformation from information symbols to TD transmit signal, the peak-to-average-power ratio (PAPR) of the transmit TD signal can be large \cite{surbhi}. Due to large PAPR, linear power amplifiers (PA) must be used to avoid signal distortion. Linear PAs are however power inefficient (the radiated power is a small fraction of the total power input to the PA) which reduces the overall energy efficiency \cite{Cripps99}.  

PAPR of OFDM modulated signals is known to be high and for which several PAPR reduction methods have been proposed in prior literature \cite{PAPR_SURVEY}. In comparison to OFDM, since OTFS is a relatively recent waveform there are some works reported on PAPR reduction for OTFS modulated signals \cite{companding, DFT, ICF}.

In this paper, we propose a novel PAPR reduction method for OTFS based systems when PSK information symbols are used. The main contributions of this paper are:
\begin{itemize}
    \item The proposed method precodes each information symbol by modifying its amplitude (without changing the phase/angle which carries information) in such a way that it helps in reducing the PAPR of the corresponding transmit TD signal (see Section \ref{prop_papr}).
    \item In Section \ref{simsec} we study the complementary cumulative distribution function (CCDF) of the PAPR, for an OTFS system where no measures are taken to reduce PAPR (i.e., referred to as an uncompensated system), for the proposed method and for the other state-of-art methods in \cite{companding, DFT, ICF}. Simulations reveal that the proposed PAPR reduction method achieves significant PAPR reduction of roughly $5$ dB when compared to an uncompensated system. Further, the proposed method has lower complexity and smaller PAPR when compared to the other state-of-art methods in \cite{companding, DFT, ICF}.
    \item In Section \ref{simsec} we also study the error rate performance of the proposed method which reveals that it achieves similar performance as that achieved with the DFT-precoded method (in \cite{DFT}) and the clipping and filtering based method (in \cite{ICF}). The proposed method achieves better error-rate performance than the companding based method (in \cite{companding}). Also, the invariance of the error rate performance of OTFS systems to a wide range of channel Doppler spread is valid even when the transmit signal is based on the proposed PAPR reduction.  
\end{itemize}      

\begin{figure*}[h]
\vspace{-4mm}
\centering
\includegraphics[width=14cm, height=2.2cm]{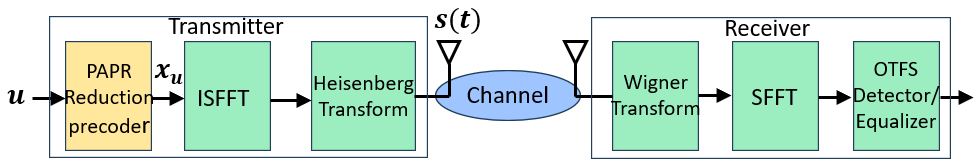}
\caption{
Zak-OTFS transceiver signal processing with proposed precoding at transmitter for PAPR reduction.}
\label{fig0}
\end{figure*}
\section{System Model}
In this paper, we consider an OTFS based wireless communication system with a single antenna base station (BS) and a single antenna user terminal (UT). In OTFS modulation, the information symbols are embedded in the delay-Doppler (DD) domain. OTFS modulation is described in detail in \cite{otfsorg}
and is parameterized by $T > 0$, and positive integers $M, N$. Let $x[k,l]$, $ k = 0,..., N- 1, l = 0,..., M-1$ denote the $MN$ DD domain information symbols.
The information symbols belong to a $D$-ary PSK alphabet set
${\mathcal S}_{A,D} = \{ A \, e^{j 2 \pi \frac{p}{D}} \, | \, p=0,1,\cdots, D-1\}$.
By using the inverse symplectic finite Fourier transform (ISFFT), the DD domain information symbols are transformed to time-frequency (TF) symbols given by 
\begin{eqnarray}\label{ISFFT}
X[n, m]&=&\sum_{k=0}^{N-1} \sum_{l=0}^{M-1} x[k, l] e^{-j 2 \pi\left(\frac{m l}{M}-\frac{n k}{N}\right)} \nonumber \\
\hspace{-5mm} & & \hspace{-10mm} n=0,1,\cdots, N-1, \,\ m=0,1,\cdots,M-1.
\end{eqnarray}



The time-frequency symbols are then converted to the transmit time domain signal $s(t)$ using the Heisenberg transform, i.e.
\begin{equation}
\label{tx_signal}
    s(t)=\sum_{n=0}^{N-1} \sum_{m=0}^{M-1} X[n, m] \, g_{t x}(t-n T) \, e^{j 2 \pi m \Delta f(t-n T)}
\end{equation}
where $g_{t x}(t)$ is the transmit pulse and $\Delta f = \frac{1}{T}$. In this paper we
consider the rectangular transmit pulse, i.e.
\begin{equation}\label{rect pulse}
    g_{tx}(t) = \begin{cases}
\frac{1}{\sqrt{T}} &, 0 \leq t < T \\
0 &, \mbox{\small{otherwise}} \\
\end{cases}.
\end{equation}The duration of $s(t)$ is $NT$ seconds and
has a bandwidth of $M \Delta f$ Hz.

\subsection{Sampling the TD transmit signal}
The transmit signal $s(t)$ is sampled at a sampling rate equal to the bandwidth $M \Delta f$ resulting in the $MN$ discrete-time samples
\begin{eqnarray}
    s[n] & \Define & s\left(t = \frac{n}{M \Delta f}\right), \,\,\,\, n=0,1,\cdots, MN -1.
\end{eqnarray}Consider the vector of discrete-time transmit samples, ${\bf s} \in {\mathbb C}^{MN}$ whose $i$-th element is $s[i-1]$, $i=1,2,\cdots, MN$. From \cite{Ravitejapulse} it follows that
\begin{eqnarray}
\label{s_vec}
{\bf s} & = & \left({\bf F}_N^H \otimes \mathbf{I}_M\right) \mathbf{\bf x}
\end{eqnarray}where ${\bf x} \in {\mathbb C}^{MN}$ is the vector of the $MN$ DD domain information symbols whose $(kM + l + 1)$-th element is $x[k,l]$, $k=0,1,\cdots, N-1,$$l=0,1,\cdots, M-1$. Further, ${\bf F}_N$ denotes the $N \times N$ DFT matrix whose element in its $p$-th row and $q$-th column is $e^{j 2 \pi p q/N}$,
$p,q=0,1,\cdots, N-1$. Also ${\bf I}_M$ denotes the $M \times M$ identity matrix and $\otimes$ denotes Kronekar product for matrices.

\section{Proposed novel approach for PAPR Compensation}
\label{prop_papr}
Peak to Average Power ratio (PAPR) is the ratio of maximum transmit power to the average transmit power in an OTFS frame and it is given by
\begin{equation}
\operatorname{PAPR}({\bf s}) \, = \, \frac{\max _{n=0, \ldots, MN-1}\left|s[n]\right|^2}{ \frac{1}{MN}\sum\limits_{n=1}^{MN} \left|s[n]\right|^2}=\frac{MN  \|\mathbf{s}\|_{\infty}^2}{ { \|\mathbf{s}\|_2^2}}.
\end{equation}

\begin{figure}[h]
\centering
\includegraphics[width=4.5cm, height=4.0cm]{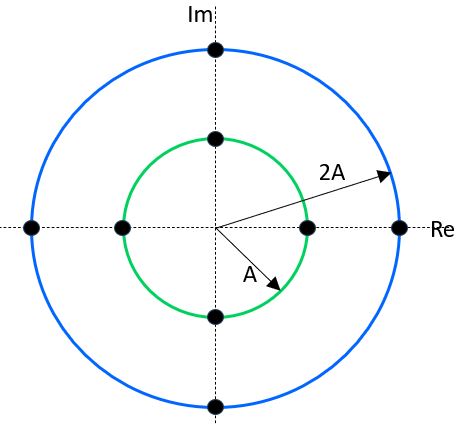}
\caption{$D=4$. Alphabet set ${\mathcal S}_{A,D}$ (black dots on the green circle) and ${\mathcal S}_{2A,D}$ (black dots on the blue circle). Precoded symbols belong to the extended alphabet set  ${\mathcal S}_{A,D} \, \bigcup {\mathcal S}_{2A,D}$ (consisting of all eight black dots on both circles). 
}
\label{fig1}
\end{figure}

In the proposed PAPR compensation method, for a given block of $MN$ information symbols , $k=0,1,\cdots, N-1$, $l=0,1,\cdots, M-1$, we precode them into precoded information symbols $x[k,l]$ which are then transmitted using (\ref{ISFFT}) and (\ref{tx_signal}). The information symbols $u[k,l]$ belong to the regular $D$-ary PSK alphabet set ${\mathcal S}_{A,D}$. However, the precoded information symbols belong to the extended alphabet set ${\mathcal S} = {\mathcal S}_{A,D} \, \bigcup {\mathcal S}_{2A,D} $. To be precise, for the $(k,l)$-th information symbol $u[k,l] \in {\mathcal S}_{A,D}$, the corresponding precoded $(k,l)$-th symbol $x[k,l]$ can be either $u[k,l]$ or $2 u[k,l]$. In PSK, information bits are encoded into the angle of the complex symbol. Since the angle of $u[k,l]$ and $2 \, u[k,l]$ is the same, information is preserved (see illustration in Fig.~\ref{fig1}).

Consider a given block of $MN$ information symbols $u[k,l]$ denoted by an information symbol vector ${\bf u} \in \left({\mathcal S}_{A,D}\right)^{MN}$. For a given ${\bf u}$ consider the set of $2^{MN}$ vectors
\begin{eqnarray}
    {\mathcal B}({\bf u}) & \hspace{-2.5mm} \Define &  \hspace{-2.5mm} {\Bigg \{}   {\bf x} \in {\mathcal S} \, {\Bigg \vert}  \, x[i] = u[i] \, \mbox{\scriptsize{or}} \, 2 u[i] \,,\, i=1,2,\cdots, MN {\Bigg \}}.
\end{eqnarray}In the proposed method we choose the precoded vector ${\bf x}$ to be the one which minimizes PAPR. For a given ${\bf u}$, we denote the chosen ${\bf x}$ by ${\bf x}_u$ which is given by
\begin{eqnarray}
\label{eqnoptprec}
    {\bf x}_u & \Define & \arg \min_{{\bf x} \in {\mathcal B}({\bf u}) }  \operatorname{PAPR}\left(  \left({\bf F}_N^H \otimes \mathbf{I}_M\right) \mathbf{\bf x} \right). 
\end{eqnarray}
The block diagram of OTFS signal processing with the proposed PAPR reduction is shown in Fig.~\ref{fig0}.
The proposed method provides a way to reduce the PAPR by allowing the precoded
symbols to take values over a larger alphabet set than that for the information symbols. However, since the number of possible precoded vectors $2^{MN}$ is large for large $(M,N)$, we next propose an iterative algorithm for approximately solving (\ref{eqnoptprec}). Although this iterative method does not solve (\ref{eqnoptprec}) exactly, it still achieves a significant PAPR improvement compared to that for the uncompensated system (see Section \ref{simsec}).

The proposed iterative method takes as input the information symbol vector ${\bf u}$ and outputs the precoded vector ${\bf x}^{\star}$ which is an approximation to the optimal ${\bf x}_u$ in (\ref{eqnoptprec}). Its algorithmic listing is provided below in Algorithm \ref{alg1list}. In step-$2$, we initialize the precoded vector to be ${\bf u}$
itself and the iteration counter $Iter$ to $0$. In each iteration (from steps $4-16$), ${\bf x}^{\star}$ denotes the best precoded vector found so far for which PAPR is $p^{\star}$. Each precoded symbol in ${\bf x}^{\star}$ has amplitude either $A$ or $2A$. In each iteration, we go through each of the $MN$ precoded symbols in ${\bf x}^{\star}$
and modify them one at a time (see steps $4-8$). In step-$6$, we scale the amplitude of the $t$-th precoded symbol by the factor $ 2^{\left( 3 - 2 \frac{\vert x^{\star}[t] \vert}{A}\right)}$ (note that we only modify the amplitude and not the angle/phase of the symbol as it carries information). It is clear that if  $\vert x^{\star}[t] \vert = A$ then the scaling factor is $2$ (i.e., the modified amplitude is $2A$) and if $\vert x^{\star}[t] \vert = 2A$ then the scaling factor is $1/2$ (i.e., the modified amplitude is $A$). The PAPR with the modified precoded vector (where only one symbol is modified) is computed in step-$7$ and saved. In step-$9$, we search for the symbol (out of all $MN$ symbols) whose modification gives the  smallest PAPR. If this smallest PAPR $p_{min}$ is less than the smallest achieved PAPR so far (i.e., $p^{\star}$), then we update $p^{\star}$ to $p_{min}$ and we update the symbol in ${\bf x}^{\star}$ whose modification resulted in the smallest PAPR so far (see steps $11-12$). However, if the smallest PAPR $p_{min}$ is not smaller than $p^{\star}$ then we terminate. We also terminate the algorithm if the maximum number of iterations $MaxIter$ is reached.

			\begin{algorithm}
			\caption{\small{Proposed Iterative Method for PAPR Compensation}}
			\label{alg1list}
				{\small
			\begin{algorithmic}[1]
				\STATE \textbf{Input:} Information symbol vector ${\bf u}$, alphabet sets ${\mathcal S}_{A,D}$,${\mathcal S}_{2A,D}$, $stop = 0$, $MaxIter = 5$.   
				
				\STATE \textbf{Initialization:} ${\bf x}^{\star} = {\bf u}$.
                $p^{\star} = \operatorname{PAPR}\left(  \left({\bf F}_N^H \otimes \mathbf{I}_M\right) {\bf u} \right)$, $Iter = 0$. 
				
				\REPEAT

                \FOR{$t = 1 \,\, to \,\, MN$} 
    
				\STATE ${\bf x} = {\bf x}^{\star}$.

                \STATE $ x[t] = 2^{\left( 3 - 2 \frac{\vert x^{\star}[t] \vert}{A}\right)} \, x^{\star}[t] $.

                \STATE $p[t] = \operatorname{PAPR}\left(  \left({\bf F}_N^H \otimes \mathbf{I}_M\right) {\bf x} \right)$.

				\ENDFOR

                \STATE $p_{min} = \hspace{-3mm} \min\limits_{t=1,2,\cdots,MN}p[t]$, $\,\,\, i = \arg \hspace{-3mm} \min\limits_{t=1,2,\cdots,MN}p[t]$.

                 \IF{$p_{min} < p^{\star}$}  
                \STATE $ x^{\star}[i] = 2^{\left( 3 - 2 \frac{\vert x^{\star}[i] \vert}{A}\right)} \, x^{\star}[i]$.
                \STATE $p^{\star} = p_{min}$.
                \ELSE
                \STATE $ stop = 1$.
                \ENDIF 
                 \STATE $Iter = Iter + 1$.
				\UNTIL{($stop = 1$ or $\,\, Iter = MaxIter$)} , \textbf{Output: ${\bf x}^{\star}, p^{\star}$.} 
				
			\end{algorithmic}}
		\end{algorithm}
For each precoded vector ${\bf x}$, the complexity of computing the transmit time-domain vector ${\bf s} =  \left({\bf F}_N^H \otimes \mathbf{I}_M\right) {\bf x}$ is $O(MN \log (MN))$. The complexity of step-$7$ is therefore $O(MN \log(MN))$. In steps $4$-$8$, the PAPR is computed for all vectors obtained
by modifying one symbol, and hence the complexity is $O(M^2 N^2 \log(MN))$.
Since the algorithm usually stops after a few iterations (less than five),
the overall complexity is $O(M^2 N^2 \log(MN))$. Other simple PAPR reduction methods like those based on companding have complexity same as that of computing the transmit time-domain vector i.e., $O(MN \log(MN))$ which is smaller than the $O(M^2 N^2 \log(MN))$ complexity of the proposed method. However, as we shall see in Section \ref{simsec}, the PAPR reduction achieved with the proposed method is significantly \emph{better} than that achieved with companding or other state-of-art methods.

\section{Numerical Simulations}
\label{simsec}
In this section, we report the simulation results for the proposed iterative PAPR compensation method, in terms of the BER performance and the complementary cumulative distribution function (CCDF) of the time-domain transmit symbols $s[n], n=1,2,\cdots, MN$.
We also compare the CCDF and BER performance of the proposed method with that of other methods used for PAPR reduction in OTFS based systems, like companding in \cite{companding}, DFT precoded OTFS in \cite{DFT}, and clipping and filtering based compensation in \cite{ICF}. For numerically computing the CCDF we compute the PAPR for a thousand different randomly generated information vectors. In the CCDF plot, for each value $x$ on the X-axis, the corresponding value on the Y-axis gives the empirically computed probability (using the thousand PAPR values) that the PAPR is greater than $x$.   

The receiver processing is same as that in \cite{otfsorg}, i.e., Wigner transform of the received time-domain signal followed by Symplectic Finite Fourier Transform (SFFT) resulting in received DD domain symbols. The received DD domain symbols can be expressed as the sum of a DD domain noise vector and the signal vector which is the product of an effective DD domain channel matrix and the vector of precoded symbols ${\bf x}$ \cite{SKM_EST}. Minimum Mean Square Error Estimation (MMSE) based equalization gives an estimate of each of the $MN$ precoded symbol. The information bits modulated onto each information symbol are then estimated from the angle/phase of the corresponding estimated precoded symbol at the output of the MMSE equalizer.       

We consider OTFS based systems with $\Delta f=1 / T=15  \mathrm{KHz}, M=16, N=16$ and a carrier frequency of $f_ c=2  \mathrm{GHz}$. The 3GPP Extended Typical Urban (ETU $300$) channel model is considered, with a maximum Doppler shift of $\nu_{\max }=300 \mathrm{~Hz}$. There are nine channel paths and the delay profile is $[0,50,120,200,230,500,1600, 2300,5000]$ ns, while the relative power profile is ${\bf P} = [-1,-1,-1,0,0,0,-3,-5,-7]$ $\mathrm{dB}$. The channel path gains $h_i, i=1,2,\cdots, 9$ are modeled as i.i.d. zero mean Rayleigh random variables having variance whose ratio is as per the power profile (i.e., ${\mathbb E}\left[ \vert h_i \vert^2 \right]/{\mathbb E}\left[ \vert h_4 \vert^2 \right] = 10^{\frac{P}{10}}$). The variance of the channel path gains are normalized so that the total variance of the channel gains is unity (i.e.,$\sum\limits_{i=1}^9 {\mathbb E}\left[ \vert h_i \vert^2 \right] = 1$). For the $i$-th channel path, the Doppler shift is modeled as $\nu_{i}=\nu_{\max } \cos \left(\theta_{i}\right)$ where $\theta_{i}, i=1,2,\cdots, 9$ are i.i.d. uniformly distributed in $[0,2 \pi)$.

\begin{figure}
    \centering    \includegraphics[width=0.85\linewidth]{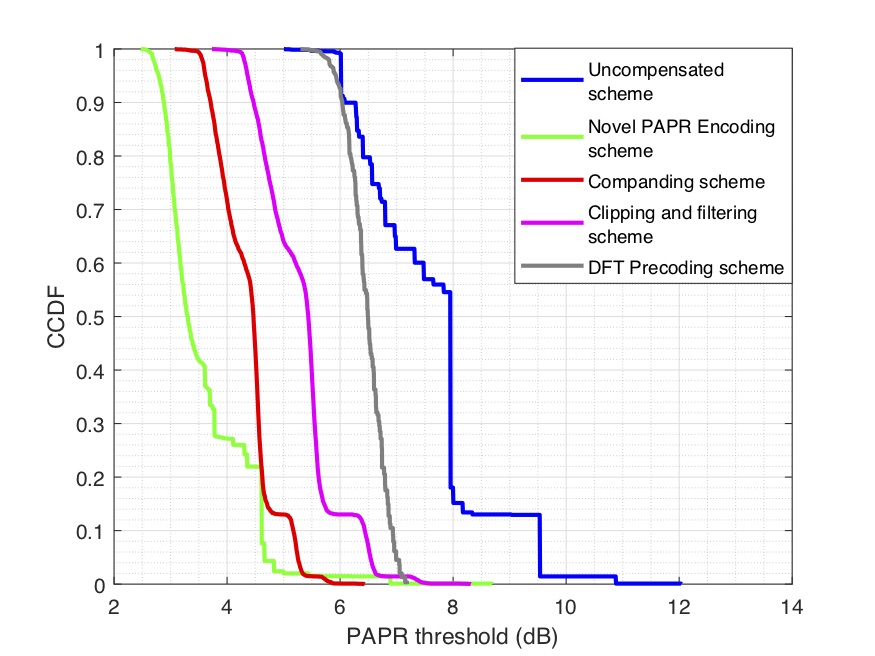}
    \caption{PAPR CCDF with BPSK.}
    \label{fig: CCDF performance}
\end{figure}

In Fig.~\ref{fig: CCDF performance} we plot the PAPR CCDF for the proposed iterative method (``Novel PAPR Encoding scheme", depicted by green curve) and that of other state-of-art methods for BPSK modulation (i.e., $D=2$ and ${\mathcal S}_{A,D} = \{A, -A \}$).
It is observed that in the absence of any compensation, the PAPR of an OTFS modulated signal (see the blue curve) can be high as also reported in \cite{surbhi}($8$ dB for at least half of the random information vectors, i.e., for CCDF = $0.5$). However, with the proposed method for $CCDF = 0.5$, the PAPR is only around $3.3$ dB as compared to $4.5$ dB for the companding based method, $5.5$ dB for the clipping and filtering based method and $6.5$ dB for the DFT-precoding based method.   

\begin{figure}
    \centering
    \includegraphics[width=0.85\linewidth]{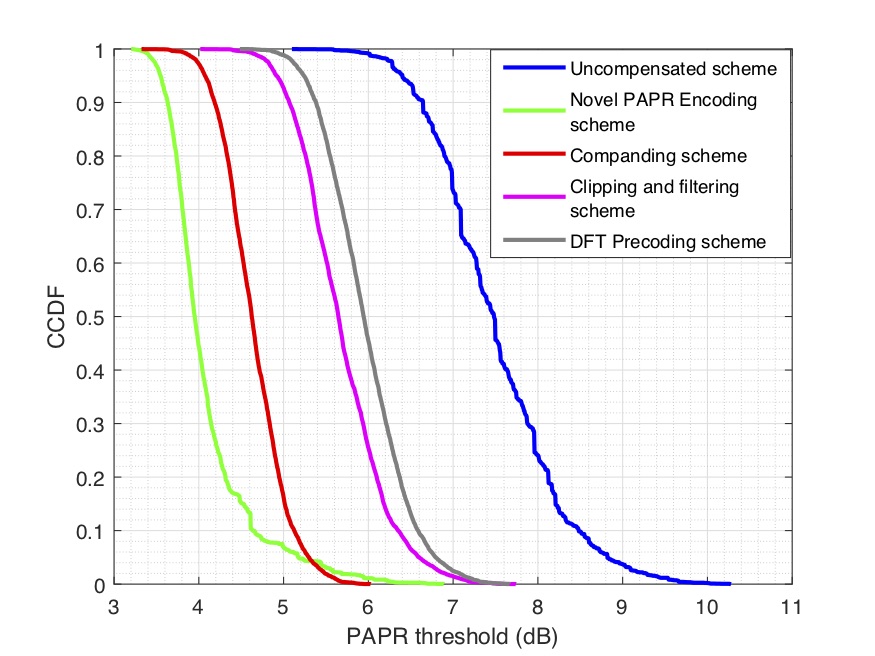}
    \caption{PAPR CCDF with QPSK.}
    \label{fig: CCDF performance of QPSK}
\end{figure}

In Fig.~\ref{fig: CCDF performance of QPSK} we plot the PAPR CCDF for OTFS modulation with QPSK alphabet (i.e., $D=4$ and ${\mathcal S}_{A,D} = \{A, A\sqrt{-1},  -A, -A\sqrt{-1} \}$). With QPSK also (just as with BPSK), the
PAPR performance of the proposed iterative method is significantly better than that of the other considered methods.

In the proposed PAPR compensation method, the average transmitted power is expected to be higher than that for the uncompensated OTFS transmission, primarily because the precoded symbols may have higher amplitude than that of the information symbols. In the following we therefore study the symbol error rate (SER) performance of the proposed compensation method in comparison with that of uncompensated OTFS and also other considered state-of-art methods. 

\begin{figure}
    \centering
    \includegraphics[width=0.85\linewidth]{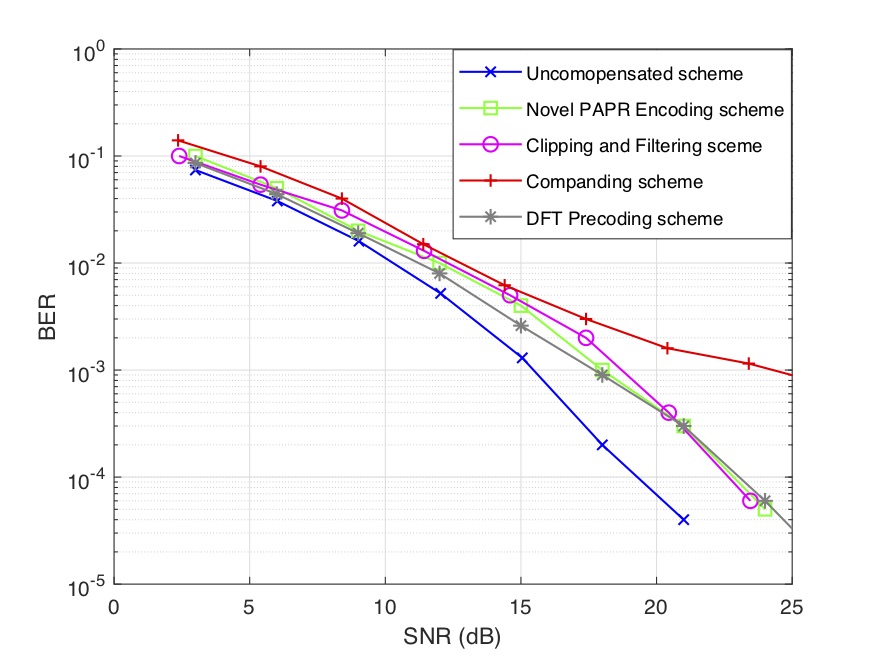}
    \caption{BER vs. SNR. BPSK information symbols.}
    \label{fig: BER performance}
\end{figure}

\begin{figure}
    \centering
    \includegraphics[width=0.85\linewidth]{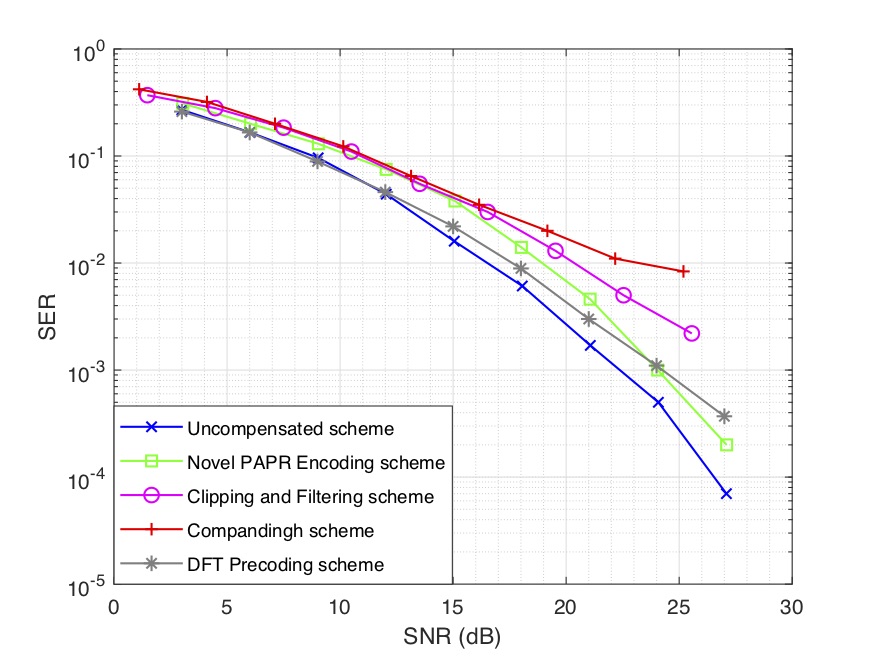}
    \caption{SER vs. SNR. QPSK information symbols.}
    \label{fig: BER performance_QPSK}
    \vspace{-3mm}
\end{figure}

In Fig.~\ref{fig: BER performance} we plot the SER/BER (bit error rate) vs. SNR performance
for all methods for OTFS modulation ($M=N=16$, $\Delta f = 15$ kHz) with BPSK information symbols. The signal-to-noise ratio (SNR) is given by the ratio of the power of the received time-domain signal to the average AWGN power at the receiver (i.e., bandwidth $M \Delta f$ times the AWGN power spectral density). From the figure it is observed that indeed the error rate performance with the proposed PAPR compensation method is slightly inferior to that of uncompensated OTFS modulation. Nevertheless, the error rate performance of he proposed compensation method is similar to that of the DFT precoded and the clipping and filtering based method and is significantly better than that of the companding based method. This fact also holds when the QPSK information symbols are used (see Fig.~\ref{fig: BER performance_QPSK}).    

\begin{figure}
    \centering
    \includegraphics[width=0.85\linewidth]{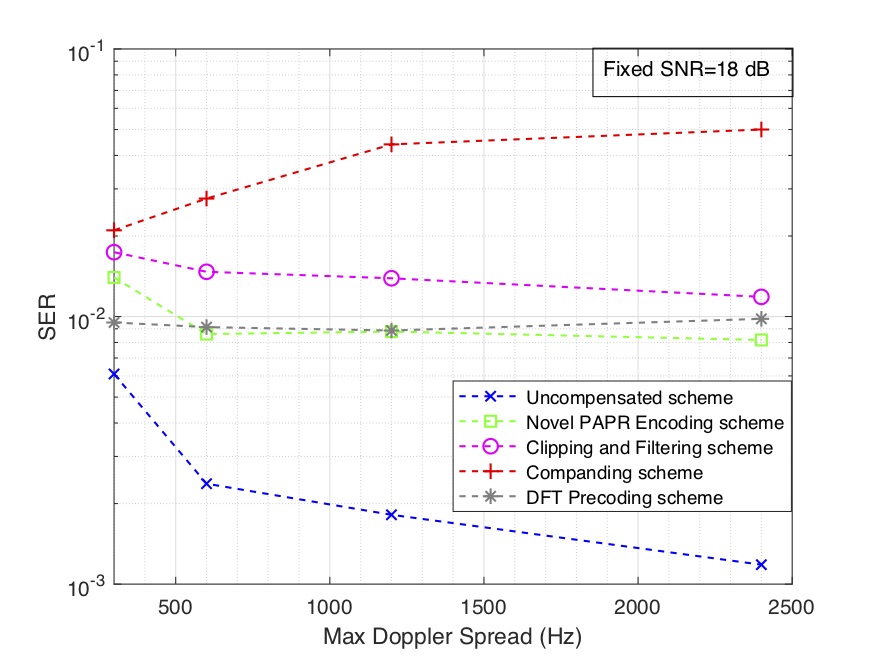}
    \caption{QPSK SER vs. max. Doppler shift $\nu_{max}$ for fixed SNR of $18$ dB.}
    \label{fig: numaxSER}
\end{figure}
 In Fig. \ref{fig: numaxSER}, we plot the QPSK SER for the proposed method and other considered state-of-art methods as a function of increasing path Doppler shift $\nu_{max}$ and a fixed SNR of $18$ dB.
It is observed that the SER performance of the proposed compensation method, the DFT precoded and the clipping and filtering based methods is almost invariant of Doppler shift ($0 \leq \nu_{max} \leq 2400$ Hz) whereas the SER performance of the companding based method degrades with increasing path Doppler shift.

In Table-\ref{tab_13} we compare the PAPR of all considered methods for
CCDF of $0.1$ with QPSK symbols, fixed $M=16$, and varying $N=4,8,16,32,64$. It is observed that the proposed method has the smallest PAPR when compared to other methods, for all values of $N$.
The same is valid for fixed $N=16$ and varying $M=4,8,16,32,64$ (see Table-\ref{tab_23}).
      \begin{table}[]
            \centering
            \caption{PAPR (dB) for CCDF $=0.1$, $M=16$, QPSK}
            \scriptsize{
            \begin{tabular}{|c|c|c|c|c|c|}
                \hline
                $N$       &  \scriptsize{Uncompen-} &  Proposed   &  Companding  &  Clipping and  & DFT  \\
                        &  sated scheme        &  encoding     & scheme   &  Filtering  &  Precoding  \\  
                \hline
                $4$   &  $6.0$ & $2.8$  & $4.1$  & $4.8$ & $6.2$ \\
                \hline
                $8$   &  $7.8$ & $2.9$  & $4.8$  & $5.9$ & $6.4$ \\
                \hline
                $16$   &  $8.5$ & $3.3$  & $5.2$  & $6.4$ & $6.6$ \\
                \hline
                $32$   &  $9.0$ & $4.0$  & $5.4$  & $6.7$ & $6.8$ \\
                \hline
                 $64$   &  $9.6$ & $4.6$  & $5.6$  & $7.2$ & $7.0$ \\
                \hline
            \end{tabular}}\normalsize
            \label{tab_13}
            \vspace{2mm}
            
            \vspace{-4mm}
        \end{table}

      \begin{table}[]
            \centering
            \caption{PAPR (dB) for CCDF $=0.1$, $N=16$, QPSK}
            \scriptsize{
            \begin{tabular}{|c|c|c|c|c|c|}
                \hline
                $M$       &  \scriptsize{Uncompen-} &  Proposed   &  Companding  &  Clipping and  & DFT  \\
                        &  sated scheme        &  encoding     & scheme   &  Filtering  &  Precoding  \\  
                \hline
                $4$   &  $7.8$ & $2.9$  & $4.9$  & $6.0$ & $4.6$ \\
                \hline
                $8$   &  $8.2$ & $3.1$  & $5.0$  & $6.2$ & $6.2$ \\
                \hline
                $16$   &  $8.4$ & $3.3$  & $5.1$  & $6.3$ & $6.7$ \\
                \hline
                $32$   &  $8.8$ & $4.6$  & $5.3$  & $6.5$ & $6.9$ \\
                \hline
                 $64$   &  $9.1$ & $5.5$  & $5.5$  & $6.8$ & $7.1$ \\
                \hline
            \end{tabular}}\normalsize
            \label{tab_23}
            \vspace{2mm}
            
            \vspace{-4mm}
        \end{table}

\section{Conclusion}
In this paper, we have proposed a novel PAPR reduction method for OTFS modulation based systems. Numerical simulations reveal that the proposed method results in significant PAPR reduction when compared to an uncompensated OTFS system. Also, the proposed method achieves lower PAPR and has lower compensation complexity than that of the other state-of-art PAPR reduction methods proposed in \cite{companding, DFT,ICF}.
The error rate performance of OTFS systems with the proposed PAPR reduction method is similar to that of other considered state-of-art methods and is invariant for a large range of channel Doppler shifts.

\end{document}